		\documentclass{elsart}
		\def\be{\begin{equation}}
		\def\ee{\end{equation}}
		\def\bea{\begin{eqnarray}}
		
		\newcommand{\newc}{\newcommand} 
		\newc{\lra}{\leftrightarrow} 
		\newc{\beq}{\begin{equation}} 
		\newc{\eeq}{\end{equation}} 
		\newc{\barr}{\begin{eqnarray}} 
		\newc{\earr}{\end{eqnarray}} 
\begin{document}
\begin{frontmatter}
		\title{The Neutrinoless Double Beta Decay from a Modern Perspective}

		\author{ 
		J.D. Vergados} 
		\address{
		Theoretical Physics Division, University of Ioannina,
		GR--451 10, Ioannina, Greece
		}
		\address{and}
		\address{
		 Department of Physics, University of Cyprus,
		P.O box 20537, CY 1878 Nicosia, Cyprus.\\
		Electronic address:Vergados@cc.uoi.gr}

		\date{\today}
		%%%%%%%%%%%%%%%%%%%%%%%%%%%%%%%%%%%%%%%%%%%%%%%%%%%%%%%%%%%%%%%%%%%%%%%
		%                          ABSTRACT                                   %
		%%%%%%%%%%%%%%%%%%%%%%%%%%%%%%%%%%%%%%%%%%%%%%%%%%%%%%%%%%%%%%%%%%%%%%%

		\begin{abstract}
		Neutrinoless double beta decay is a very important process 
		both from the particle and nuclear physics point of view. 
		From the elementary particle point of view it pops up in almost
		every model, giving rise, among others, to the following mechanisms: 
		a) The traditional contributions
		like the light neutrino mass mechanism as well as the $j_{L}-j_R$ leptonic
		interference ($\lambda$ and $\eta$ terms). b)  The exotic R-parity violating
		supersymmetric (SUSY) contributions. In this scheme the
                currents are left handed only and the intermediate particles
               normally are very heavy. There exists, however, the possibility
               of light intermediate neutrinos arising by the combination of
               V-A and P-S currents at the quark level. This leads to the 
               same structure as the above $\lambda$ term. Similar 
               considerations apply to its sister lepton and muon number 
               violationg muon to positron conversion in the presence of nuclei.

               Anyway, regardless
               of the dominant mechanism, the observation of neutrinoless
               double betas decay, which is the most important of the two from
               an experimental point of view,  
		will severely constrain the existing models and 
		will signal that the neutrinos are massive Majorana particles.
		From the nuclear
		physics point of view it is challenging, because: 1) The nuclei, which can 
		undergo double beta decay, have complicated nuclear structure. 2) The
		energetically allowed transitions are suppressed (exhaust a small part of all
		the strength). 3) Since in some mechanisms the intermediate particles are very
		heavy, one must
		cope with the short distance behavior of the transition operators. Thus novel 
		effects, like  the double beta decay of pions in flight between nucleons, 
		have to be considered. In SUSY models this mechanism is more important than
		the standard two nucleon mechanism.  
		4) The intermediate momenta involved are quite high (about 100 $MeV/c$). Thus 
		one has to take into account  possible momentum 
		dependent terms  of the nucleon current, like the modification of the axial
                current due to PCAC, weak magnetism terms etc. 
		We find that, for the mass mechanism, 
		such modifications of the
		nucleon current for light neutrinos reduce the nuclear matrix elements by about
		$25\%$, almost regardless of the nuclear model.
		In the case of heavy neutrino the effect is much larger and model dependent.
		 Taking the above effects into account the needed nuclear matrix elements 
		have become available  for all the experimentally  interesting nuclei A = 76,
		82, 96, 100, 116, 128, 130, 136 and 150. Some of them have been obtained in the
                large basis shell
                model but most of them in various versions of QRPA. 
		 Then  using the best presently available experimental limits on the
		half-life of the $0\nu\beta\beta$-decay
		, we have extracted new limits on the  various lepton violating parameters. 
		In particular we find $\langle m_{\nu} \rangle < 0.5eV/c^2$ and, for reasonable 
		choices of the parameters of SUSY
               models in the allowed SUSY parameter space, we
		get a stringent limit on the R-parity violating parameter 
		$\lambda^{\prime}_{111} < 0.68 \times 10^{-3}$
		\end{abstract}
		\begin{keyword}
		PACS numbers:23.40.Hc,21.60.Jz,27.50.+e,27.60.+j\\\\\
		\end{keyword}
\end{frontmatter}
		%%%%%%%%%%%%%%%%%%%%%%%%%%%%%%%%%%%%%%%%%%%%%%%%%%%%%%%%%%%%%%%%%%%%%%%
		%                          INTRODUCTION                               %
%%%%%%%%%%%%%%%%%%%%%%%%%%%%%%%%%%%%%%%%%%%%%%%%%%%%%%%%%%%%%%%%%%%%%%%
%\newpage
% \vspace{5.0cm}
%{\bf Contents}
\section{Introduction}
\section{The intermediate Majorana neutrino mechanism}
\subsection{The Majorana neutrino mass mechanism}
\subsection{The leptronic left- right interference term ($\lambda$ and
$ \eta$ terms}
\subsection{The Majoron emission mechanism}
\section{Brief description of the current experiments}
\section{The R-parity violating contribution to $0\nu \beta \beta$-decay}
\subsection{The contribution arising from the bilinears in the superpotential}
\subsection{The contribution arising from the cubic terms in the
 superpotential}
\subsection{The case of light intermediate neutrinos}
\section{The effective nucleon current}
\subsection{Handling the short range nature of the transition operator}
\subsubsection{The momentum-dependent corrections to the effective nucleon
current}
\section{The exotic double charge exchange to $\mu^-$ to $e^+$ conversion
 in nuclei}
\subsection{The transition operator at the nuclear level}
\subsection{Irreducible tensor operators}
\subsection{The branching ratio for $(\mu^-,e^+)$}
\subsection{Results and discussion for $(\mu^-,e^+)$ conversion}
\subsection{Summary and conclusions}
\section{Extraction of the lepton violating parameters}
\subsection{Traditional lepton violating parameters}
\subsection{R-parity induced lepton violating parameters}
\section{Conclusions}
\section{Acknowledgments}
\section{References}
\end{document}